\documentclass[12pt,preprint]{aastex}
\usepackage{natbib}

\newcommand{\kms}{\ifmmode {\rm km\ s}^{-1} \else km s$^{-1}$\fi}
\newcommand{\Msun}{\ifmmode {\rm M}_{\odot} \else M$_{\odot}$\fi}
\newcommand{\Lsun}{\ifmmode {\rm L}_{\odot} \else L$_{\odot}$\fi}
\newcommand{\qo}{\ifmmode q_{\rm o} \else $q_{\rm o}$\fi}
\newcommand{\Ho}{\ifmmode H_{\rm o} \else $H_{\rm o}$\fi}
\newcommand{\ho}{\ifmmode h_{\rm o} \else $h_{\rm o}$\fi}

\newcommand{\vFWHM}{\ifmmode v_{\mbox{\tiny FWHM}} \else
                    $v_{\mbox{\tiny FWHM}}$\fi}
\newcommand{\CCF}{\ifmmode F_{\it CCF} \else $F_{\it CCF}$\fi}
\newcommand{\ACF}{\ifmmode F_{\it ACF} \else $F_{\it ACF}$\fi}
\newcommand{\Halpha}{\ifmmode {\rm H}\alpha \else H$\alpha$\fi}
\newcommand{\Hbeta}{\ifmmode {\rm H}\beta \else H$\beta$\fi}
\newcommand{\Hgamma}{\ifmmode {\rm H}\gamma \else H$\gamma$\fi}
\newcommand{\Hdelta}{\ifmmode {\rm H}\delta \else H$\delta$\fi}
\newcommand{\Lya}{\ifmmode {\rm Ly}\alpha \else Ly$\alpha$\fi}
\newcommand{\Lyb}{\ifmmode {\rm Ly}\beta \else Ly$\beta$\fi}
\newcommand{\HeI}{\ifmmode {\rm He}\,{\sc i}\,\lambda5876 \else 
	          He\,{\sc i}\,$\lambda5876$\fi}
\newcommand{\HeII}{\ifmmode {\rm He}\,{\sc ii}\,\lambda4686 \else 
	           He\,{\sc ii}\,$\lambda4686$\fi}

\newcommand{\feii}{Fe\,{\sc ii}}

\newcommand{\ciii}{\ifmmode {\rm C}\,{\sc iii} \else C\,{\sc iii}\fi}
\newcommand{\civ}{\ifmmode {\rm C}\,{\sc iv} \else C\,{\sc iv}\fi}

\newcommand{\niv}{N\,{\sc iv}}

\newcommand{\oiii}{O\,{\sc iii}}
\newcommand{\ob}{[O\,{\sc iii}]\,$\lambda 4959,5007$}

\newcommand{\mgii}{Mg\,{\sc ii}}

\slugcomment{To be published in Astrophysical Journal}

\shorttitle{Quasar Black Hole Masses at High Redshifts}
\shortauthors{Dietrich et al.}

\begin{document}

%
%

\title{Implications of Quasar Black Hole Masses at High Redshifts}

\author{
M.\,Dietrich
 \altaffilmark{1,2}
and
F.\,Hamann
 \altaffilmark{2},
}
\altaffiltext{1}
{Department of Physics and Astronomy, Georgia State University, 
 One Park Place South SE, Atlanta, GA 30303, USA.}
\altaffiltext{2}
{Department of Astronomy, University of Florida, 211 Bryant Space Science 
 Center, Gainesville, FL 32611-2055, USA.}   
\email{dietrich@chara.gsu.edu}

\begin{abstract}
We investigated a sample of 15 luminous high-redshift quasars
($3.3 \la z \la 5.1$) to measure the mass of their super-massive black holes 
(SMBH) and compare, for the first time, results based on \civ , \mgii , and 
H$\beta$ emission lines at high-redshifts.
Assuming gravitationally bound orbits as dominant broad-line region gas motion,
we determine black hole masses in the range of $M_{bh}\simeq 2 \times 10^8 
M_\sun$ up to $M_{bh}\simeq 4 \times 10^{10} M_\sun$.
While the black hole mass estimates based on \civ\ and H$\beta $ agree well,
\mgii\ typically indicates a factor of $\sim 5 \times$ lower SMBH masses. 
     A flatter slope of the H$\beta$ radius -- luminosity relation, a possibly 
     steeper slope of the \mgii\ radius -- luminosity relation, and a slightly
     larger radius of the \mgii\ BLR than for H$\beta$ could relax the 
     discrepancy.
     In spite of these uncertainties, the \civ , \mgii , and H$\beta$ emission 
     lines consistently indicate super-massive black hole masses of several
     times $10^9 M_\sun$ at redshifts up to z=5.1.
Assuming logarithmic growth by spherical accretion with a mass to energy 
conversion efficiency of $\epsilon = 0.1$ and an Eddington ratio 
$L_{bol}/L_{edd}$ calculated for each quasar individually, 
we estimate black hole growth-times of the order of several $\sim 100$ Myr 
which are smaller than the age of the universe at the corresponding redshift.
Assuming high-mass seed black holes ($M_{bh}^{seed} = 10^3 \,{\rm to}\, 10^5 
M_\odot$) the SMBHs in the $z\simeq 3.5$ quasars began to grow at redshifts 
$z \ga 4$, while for the quasars with  $z\ga 4.5$ they started at 
$z \simeq 6 \,{\rm to}\, 10$. 
These estimated time scales for forming SMBHs at high redshifts, together with 
previous studies indicating high quasar metallicities, suggest that the main 
SMBH growth phase occurs roughly contemporaneously with a period of violent and
extensive star formation in proto-galactic nuclei.

\end{abstract}
\section{Introduction}
During the last few years it has been established that almost all
massive galaxies are hosting a super-massive black hole (SMBH) in their 
centers. It has been found that the SMHB mass is well correlated with the 
host galaxy's spheroidal component mass and with the bulge luminosity 
(e.g., Kormendy
\& Richstone 1995; Magorrian et al.\,1998; McLure \& Dunlop 2001). 
Furthermore, a remarkably close correlation of the SMBH mass and the stellar 
velocity dispersion, $\sigma _\ast$, is detected for active, as well as for 
non-active galaxies (Gebhardt et al.\,2000; Graham et al.\,2001;
Merritt \& Ferrarese 2001; McLure \& Dunlop 2002; Tremaine et al.\,2002).
These results, together with indications that luminous quasars reside in 
massive early type galaxies (e.g., McLure et al.\,1999; Kukula et al.\,2001;
Nolan et al.\,2001; Percival et al.\,2001; Dunlop et al.\,2003), point to a 
closely related formation of massive spheroidal galaxies and the formation 
and growth of central SMBH.
However, it is still under debate whether the black hole and the host galaxy 
formed simultaneously or whether one component formed before the other 
(e.g., Silk \& Rees 1998; Cattaneo et al.\,1999; Salucci et al.\,1999; 
Kauffmann \& Haehnelt 2000; Haiman \& Loeb 2001; Menou et al.\,2001;
Archibald et al.\,2002; Di\,Matteo et al.\,2003).

It is of great importance to estimate SMBH masses at higher redshifts to 
investigate whether the relations of SMBH mass with fundamental parameters of 
their harboring galaxies that are found for active and non-active galaxies in 
the local universe are already established at early cosmic epochs.
Recent studies of quasars at moderate (McLure \& Jarvis 2002, $z\la 1$) and 
high-redshifts (Shields et al.\,2003, $z\la 3$; Vestergaard 2004, $z\la 6$) 
assume that the locally determined radius -- luminosity relation can be 
applied up to redshifts of $z\simeq 6$.
Particularly, black hole masses have been estimated for several high-redshift
quasars, providing evidence that even at early cosmic times black holes with 
several times $10^9 M_\odot$ are already in place (Barth et al.\,2003; 
Willott et al.\,2003; Vestergaard 2004; Warner et al.\,2004).
The detection of quasars at redshifts $z\ga 6$ (Fan et al.\,2001)
and black hole mass estimates well beyond ${\rm M}_{bh}\simeq 10^{10} M_\odot$ 
(Bechthold et al.\,2003; Netzer 2003) challenges black hole growth models
to buildup SMBHs on time scales of several times $\sim 100$ Myr, because
at $z\ga6$ the age of the universe is less than 0.9 Gyr.
However, the high-redshift $M_{bh}$ estimates are all based on the 
\civ $\lambda 1549$ emission line and the assumption of virialized gas motion. 
It is obviously important to use more than a single emission line to determine
$M_{bh}$ as a fundamental AGN parameter.

Much effort has been spent to derive methods to estimate the black hole
mass, $M_{bh}$, based on reverberation experiments and assuming virial
dominated motion of the broad line region (BLR) gas (e.g., Kaspi et al.\,2000; 
Peterson \& Wandel 1999,2000; Wandel, Peterson, \& Malkan 1999). This results 
in an 
empirically established radius-luminosity relation, $R_{BLR} \propto L^\beta$ 
with e.g. $\beta = 0.7$ (Kaspi et al.\,2000). Since long-term variability 
studies are currently not feasible for quasars, alternative methods are 
suggested employing single epoch emission line profile width measurements to 
estimate $M_{bh}$ (Vestergaard 2002). There is still some concern regarding 
this approach (e.g., Krolik 2001) and that the R-L relation might be flatter 
than $\beta =0.7$ (e.g., Peterson et al.\,2000; McLure \& Jarvis 2002; 
Corbett et al.\,2003; Netzer 2003; Vestergaard 2002, 2004 for a more detailed 
discussion).
However, using emission line profiles of H$\beta $ and \civ $\lambda 1549$
to estimate the SMBH mass of quasars yields consistent results that are on
average within $\sim 20$\,\%\
(e.g., Vestergaard 2002; Warner et al.\,2003). Recently, McLure \& Jarvis 
(2002) suggested \mgii $\lambda 2798$ to measure black hole masses. 
Particularly for quasars at redshifts $z \ga 4$, \mgii\ is of great interest 
because H$\beta $ is shifted out of the K-band. 
As a low ionization line \mgii\ can be used as a well suited replacement of 
H$\beta$ instead of using the high ionization line \civ $\lambda 1549$.
In addition, independent methods based on the well established 
$M_{bh} \propto \sigma _\ast$ and $M_{bh} \propto M_{bulge}$ relations for 
active and non-active galaxies (Gebhardt et al.\,2000; Ferrarese et al.\,2001; 
Laor 2001; McLure \& Dunlop 2001,2002; Tremaine et al.\,2002) result in 
consistent black hole mass estimates (see Barth 2003; Vestergaard 2004 for a 
more detailed discussion).

In the following we present the results of SMBH mass estimates for luminous
quasars at redshifts $3.3 \la z \la 5.1$ that we observed in the near infrared
to study the evolution of the \feii /\mgii\ emission line ratio 
(Dietrich et al.\,2002a,2003a).
We analyzed the broad emission line profiles of \mgii $\lambda 2798$ and  
H$\beta$. For most of the quasars we also have observed or have access to 
the ultraviolet spectral range that contains the strong \civ $\lambda 1549$ 
emission line. Hence, we could use almost always at least two emission line
profiles to determine $M_{bh}$ and for the first time to compare the obtained
SMBH masses.

In the following we assume $H_o = 65$ km\,s$^{-1}$\,Mpc$^{-1}$, 
$\Omega _M = 0.3$, $\Omega _\Lambda = 0.0$. 
Introducing $\Omega_\Lambda = 0.7$ (Netterfield et al.\,2002) instead of
$\Omega_\Lambda = 0.0$, the luminosities at the highest redshifts would be 
$\sim 10$\,\%\ larger. Hence, a slight change in luminosity will only have 
a minor impact on the estimated black hole mass.

\section{The High-Redshift Quasar Sample}
We have observed a sample of 15 luminous high-redshift quasars with 
$z \simeq 3.3$ up to $z \simeq 5.1$ (Table 1).
The spectra were obtained using infrared and optical spectrographs of the 
Cerro Tololo Inter-American Observatory at Chile (OSIRIS, 4\,m),
the ESO observatories at La Silla (SofI, 3.5\,m) and Paranal (ISAAC, 8.2\,m)
at Chile, the Calar Alto Observatory (Twin, 3.5\,m) at Spain, and the 
W.M.\,Keck Observatory (NIRSPEC, 10\,m) at Hawai'i (for more details on the 
observations see Dietrich et al.\,1999,2002a,2003a; 
 Diet\-rich \& Wilhelm-Erkens 2000).
In order to study the broad ultraviolet \feii\ emission and the 
\mgii $\lambda 2798$ emission line ratio J, H, and K-band spectra were
recorded.
For the quasars with $z\simeq 3.5$ we observed in the K-band the rest frame 
optical wavelength range which covers H$\beta$ and \ob .
Furthermore, for 10 of the studied quasars we have ultraviolet spectra of the 
\civ $\lambda 1549$ emission line (Dietrich et al.\,2002b).

\section{Data Analysis}
\subsection{Emission Line Profile Measurements}
For each of the high-redshift quasars we measured the emission line profile
width of H$\beta $, \mgii $\lambda 2798$, and \civ $\lambda 1549$ if covered.
To determine a reliable profile width it is necessary to identify and to 
remove potential contamination by additional emission features.
Although the \mgii $\lambda 2798$ emission line doublet appears to be
relatively isolated in quasar spectra, it is embedded in the superposition of 
many thousands of discrete ultraviolet \feii\ emission lines that form broad 
emission blends (e.g., Wills, Netzer, \& Wills 1985). The careful correction 
for ultraviolet Fe emission yields a more solid measurement of the \mgii\ 
profile width than assuming a linear pseudo-continuum beneath the \mgii\ 
emission line.
The contamination of H$\beta $ by optical \feii\ emission is less severe 
(e.g., Boroson \& Green 1992). \civ $\lambda 1549$ suffers only minor 
contamination of weak emission lines, like \niv ]$\lambda 1486$ in the blue 
wing and from \feii\ emission beneath the \civ\ line profile 
(e.g., Vestergaard \& Wilkes 2001).
To account for these emission line profile contaminations we analyzed the high 
redshift quasar spectra using a multi-component fit approach 
(Dietrich et al.\,2002a,2003a).
Following Wills, Netzer, \& Wills (1985) we used the following four components 
to reconstruct each quasar spectrum, 
    (i)   a power-law continuum (F$_\nu \sim \nu ^{\alpha}$),
    (ii)  Balmer continuum emission (Grandi 1982; Storey \& Hummer 1995),
    (iii) \feii\ emission templates for the ultraviolet and optical
          wavelength range (Vestergaard \& Wilkes 2001; Boroson \& Green 1992),
and (iv)  and Gaussian profiles to account for individual broad emission 
          lines, such as H$\beta$, \mgii , and \civ .
The power-law continuum fit to the individual quasar spectra was iteratively 
determined taking into account the measured strength of the \feii\ 
emission and Balmer continuum emission derived by the fitting process.
The emission line profile widths, expressed as full width at half maximum 
(FWHM) which we measured for H$\beta$, \mgii , and \civ\ if are accessible, 
are given in Table 2, together with the spectral slope $\alpha$ of the 
power-law continuum.
Since the spectral resolution of the infrared spectra is of the order of
$\sim 200$ to 500 km\,s$^{-1}$ (Table 2), a correction of the measured profile 
width with respect to the instrumental resolution is not necessary. 

\subsection{Estimating Black Hole Masses}
%
The empirically justified radius-luminosity relation, $R_{BLR} \propto L^\beta$
(e.g., $\beta = 0.7$, Kaspi et al.\,2000) and the assumption of virial 
dominated motion of the broad line region (BLR) gas (Peterson \& Wandel 1999; 
Onken \& Peterson 2002) has led to equations which connect the black hole mass
$M_{bh}$ with observationally accessible parameters like luminosity and 
emission line profile width. 
We applied the equations given by Kaspi et al.\,(2000), McLure \& Jarvis 
(2002), Vestergaard (2002), and Warner et al.\,(2003) to estimate the black 
hole mass for the high-redshift quasars we observed. 
The explicit form of the equations we employed, are presented in the appendix 
for H$\beta$, \mgii $\lambda 2798$, and \civ $\lambda 1549$.

It should be kept in mind that although \civ $\lambda 1549$ is a 
high-ionization emission line, in contrast to H$\beta$ and \mgii$\lambda 2798$
as typical low-ionization lines, there are good reasons to assume that the 
broad line \civ\ emitting gas motion also is gravitationally dominated
(Peterson \& Wandel 2000). This assumption is based on variability response, 
the emission line profile shape, and the relation of the profile widths of 
H$\beta$ and \civ $\lambda 1459$ combined with variability time lags
(Onken \& Peterson 2002).
Furthermore, it has been shown that using emission line profiles of H$\beta$ 
and of \civ $\lambda 1549$ to estimate the SMBH mass of quasars yields 
consistent results on average (e.g., Vestergaard 2002; Warner et al.\,2003). 

\section{Results}
\subsection{Super Massive Black Hole Mass Estimates}
We applied the equations (A1) to (A5) given in the appendix together with the 
emission line 
profile widths of H$\beta $, \mgii , and \civ\ and the spectral index $\alpha$ 
of the continuum slope (Table 2), to calculate the mass of the 
SMBH for each of the high-redshift quasars. 
To estimate the uncertainty of the determined SMBH masses the errors of the 
emission line profile width measurements and of the continuum luminosity have 
been propagated.  
In Table 3 the resulting black hole masses for each emission line estimator 
are presented. In Figure 1 the black hole mass is displayed for each emission 
line as a function of redshift. 
The luminous quasars under study have black hole masses in the range of 
${\rm M}_{bh}\simeq 2 \times 10^8 M_\sun$ up to ${\rm M}_{bh}\simeq 4 
\times 10^{10} M_\sun$. 
The SMBH mass based on the H$\beta $ profile width are consistent within a 
factor of $\sim 1.9\pm 0.1$, with lower mass estimates provide by the relation 
suggested by McLure \& Jarvis (2002) compared to those of Kaspi et al.\,(2000).
This difference is mainly caused by the different slopes of the applied R -- L
relation. 
     Assuming a slope of $\beta = 0.6$ instead of $\beta = 0.7$ will result in
     $\sim 2$ times lower $M_{bh}$ estimates.
     Utilizing the \civ\ emission line profile width, the resulting SMBH mass 
     estimates, using equations A3 and A4 as given in the appendix, are 
     consistent within $\sim 15$\,\% , as expected since the same 
     slope is assumed for the R -- L relation (eqs. A3 and A4).
     The scatter of the two SMBH estimates based on \civ\ are due to the use of
     the individual spectral index $\alpha$ for each quasar that we
     determined instead of using a uniform continuum slope of $\alpha = -0.4$
     for example. Furthermore, to use the continuum luminosity L(1450 \AA )
     for each quasar an additional $\alpha$-dependent factor is introduced.
     In spite of these modifications the resulting black hole mass estimates,
     using the \civ\ profile width, agree well with each other.
Close inspection of Figure 3 obviously shows that the $M_{bh}$ estimates based
on \mgii\ are systematically lower than the ones based on H$\beta$ and 
\civ $\lambda 1549$. On average the \mgii $\lambda 2798$ emission line profile 
indicates a $\sim 5 \times$ lower SMBH mass than the H$\beta $ and 
\civ $\lambda 1549$ line profile. 
For each of the high-redshift quasars we calculated a mean SMBH mass based on 
H$\beta$, \mgii , and \civ\ if available (Table 3). Although the SMBH mass 
estimates based on \mgii\ are lower than those given by H$\beta $ and \civ , 
we do not apply possible corrections (see Sect.\,4.2).

\subsection{Size of the MgII Emission Region}
A basic assumption of the R--L relation involving the \mgii $\lambda 2798$
emission line is that the emission regions of \mgii\ and H$\beta$ are nearly at
the same distance from the central continuum source. While the comparison of 
the profile width of both emission lines indicate that $R_{BLR}(H\beta)$ is
nearly identical with $R_{BLR}({\rm MgII})$ (McLure \& Jarvis 2002), 
Corbett et al.\,(2003) find that \mgii\ has generally a broader profile than
H$\beta$. 
In Figure 2 we compare the FWHM of the line profiles of \mgii\ and H$\beta$
for the seven quasars at $z\simeq 3.5$. With the exception of Q\,2050$-$359 
and Q\,0105$-$2634 (similar width) and Q\,2348$-$4025 (larger \mgii\ profile 
than H$\beta$) the H$\beta $ profile is generally slightly broader than 
the \mgii\ profile. 
     The average ratio of the FWHM of the H$\beta$ to \mgii $\lambda 2798$ line
     profile amounts to $1.2 \pm 0.5$ 
     based on the seven $z\simeq 3.5$ quasars we studied.

The most direct measure of the BLR radius for \mgii\ is given by reverberation 
mapping experiments. So far, almost all monitoring campaigns dedicated to AGNs 
have been focused in the optical on the H$\beta $ region and in the ultraviolet
on the Ly$\alpha$ -- \civ $\lambda 1549$ -- \ciii ]$\lambda 1909$ wavelength 
range. However, for NGC\,5548 (Clavel et al.\,1991; Peterson et al.\,1991; 
Dietrich \& Kollatschny 1995) and NGC\,3783 (Reichert et al.\,1994; 
Stirpe et al.\,1994) \mgii $\lambda 2798$ and H$\beta $ were measured.
NGC\,3783 showed only minor \mgii\ emission line flux variations during the 
seven month campaign in 1992. The derived emission line flux variation delays 
for \mgii\ and H$\beta$ are of the same order of $\tau \simeq 6 - 8$ days 
(Reichert et al.\,1994; Stirpe et al.\,1994).
In contrast, NGC\,5548 underwent quite strong variations for H$\beta$ and 
\mgii\ during the first year of its monitoring in 1989. Dietrich \& Kollatschny
(1995) used the NGC\,5548 observations taken with IUE in the ultraviolet 
(Clavel et al.\,1991) and from the ground-based monitoring campaign 
(Peterson et al.\,1991) to estimate a typical radius of the 
\mgii $\lambda 2798$ and H$\beta$ emission line regions. They determined time 
delays based on the centroid ($\tau _c$), as well as on the peak ($\tau _p$), 
of the interpolated cross-correlations function (ICCF). 
The time delays based on the ICCF centroid amount to 
$\tau _c {\rm (H\beta)} = 19.7 \pm 2.4$\,d and
$\tau _c {\rm (MgII)} = 46.5 \pm 2.3$\,d, while the peak of the ICCF
yield delays of
$\tau _p {\rm (H\beta)} = 19.5 \pm 2.4$\,d and
$\tau _p {\rm (MgII)} = 39.5 \pm 3.3$\,d
 (Dietrich \& Kollatschny 1995).
Assuming that these delays represent typical distances of the line emitting gas
from the central variable continuum source, in the case of NGC\,5548 the 
\mgii\ gas is located $\sim 2$ times more distant than the H$\beta$ emitting
region. In a very simplified model of gravitationally bound Keplerian orbits 
it can be expected that the typical velocity of \mgii\ is about $\sqrt{2}$ 
times narrower than for H$\beta$.
     However, this estimate of $R_{BLR}$(\mgii) is based on only one Seyfert 
     galaxy.
     Although this result is quite consistent with the ratio of the profile 
     widths of H$\beta $ to \mgii\ that we find for the high-redshift quasars 
     (Figure 2 and Table 2), it is necessary to measure $R_{BLR}$(\mgii) for 
     more Seyfert galaxies and quasars. 
Since a larger distance of the \mgii\ BLR emitting region than for H$\beta$ 
will introduce a factor of $b = R_{BLR}{\rm (MgII)} / R_{BLR}{\rm (H\beta)}$,
the estimated SMBH mass will be slightly underestimated assuming 
$R_{BLR}{\rm (MgII)} \simeq R_{BLR}{\rm (H\beta)}$.

\subsection{Bolometric Luminosity and Eddington Ratio}
To compare the SMBH mass with the bolometric luminosity, $L_{bol}$, and to 
determine the Eddington luminosity ratio, $L_{bol} / L_{edd}$, we 
calculated $L_{bol}$. To transform measured rest frame luminosities 
$\lambda L_\lambda$ into $L_{bol}$ scaling factors are provided by e.g. 
Elvis et al.\,(1994) and Kaspi et al.\,(2000). 
Most recently, Vestergaard (2004) and Warner et al.\,(2004) present revised 
scaling 
factors for optical and ultraviolet continuum luminosities employing results of
multi-frequency studies on the AGN spectral energy distributions from X-ray 
energies to the mid-infrared wavelength range.
They found consistent scaling factors which are slightly smaller 
($\sim 20$\,\%) than previously assumed. We used the relation
$L_{bol} = 4.36 \times \lambda L_\lambda (1450 {\rm \AA})$ to calculated the 
bolometric luminosity (Warner et al.\,2004) for each of the high-redshift 
quasars under study (Table 1).
In Figure 3 we plot the average SMBH masses as a function of L$_{bol}$. 
At a given L$_{bol}$ the determined M$_{bh}$ of the high-redshift quasars 
cover a range of up to one order of magnitude in mass, similar to comparable 
studies for quasars at lower redshift (e.g., Woo \& Urry 2003; 
Vestergaard 2004; Warner et al.\,2004).

For a fully ionized pure hydrogen gas, i.e. $\mu_ e = 1$, the Eddington 
luminosity is given as (Rees 1984; Peterson 1997)

\begin{equation}
L_{edd} = 1.26 \times 10^{38} \, \biggl({M_{bh} \over M_\sun}\biggr) \,
          \mu_e \,\,\,{\rm erg \,\,s}^{-1}
\end{equation}

Assuming a more realistic gas chemical composition containing hydrogen, 
helium, and some heavier elements will result in an additional factor 
$\mu_e = 1.15$ accounting for the atomic weight per electron (Haiman \& Loeb 
2001). This will yield a 15\,\%\ higher Eddington luminosity. 
In the following we will assume pure hydrogen gas because $\mu \neq 1$ has 
only minor impact on $L_{edd}$.

The Eddington ratio $L_{bol}/L_{edd}$ is determined for each quasar 
individually. 
The mean black hole masses are used to calculate the Eddington luminosity, 
$L_{edd}$, for spherical accretion. Together with the bolometric luminosity, 
the Eddington ratio, $L_{bol} / L_{edd}$, is calculated (Table 3).
The determined $L_{bol} / L_{edd}$ covers a range of $\sim 0.5$ to $\sim 7$
which is in good agreement with results of studies of quasars sample that are
dominated by low and intermediate redshift quasars 
(e.g., Woo \& Urry 2003; Vestergaard 2004; Warner et al.\,2004).
Hence, there is no difference regarding the Eddington ratio $L_{bol} / L_{edd}$
for quasars at high-redshifts compared to those at lower redshift.

\section{Discussion}

\subsection{Comparison of $M_{bh}$ based on H$\beta$, \mgii , and \civ }

In Table 3 the SMBH mass estimates based on the three emission lines
H$\beta$, \mgii $\lambda 2798$, and \civ $\lambda 1549$ are listed for 
comparison.
Similar to previous studies there appears to be an approximate upper limit for 
$M_{bh}$ at $\sim 10^{10} M_\odot$ (e.g., Corbett et al.\,2003; Netzer 2003;
Shields et al.\,2003; Vestergaard 2004; Woo \& Urry 2003; Warner et al.\,2003).
Only Q\,0105-2634 yields an estimate of $M_{bh} = (4.1 \pm 1.2) \times 
10^{10} M_\odot$ (eq.\,A1) and $(2.1 \pm 0.4) \times 10^{10} M_\odot$ 
(eq.\,A2), 
respectively, based on the H$\beta$ emission line width. However, the H$\beta$
emission profile of this particular quasar has a rather low signal-to-noise 
ratio. Furthermore, the estimate $M_{bh} = (4.1 \pm 1.2) \times 10^{10} 
M_\odot$ based on the R--L relation with a slope of $\beta = 0.7$. Assuming a 
flatter slope $\beta$ of the R--L relation, as suggested by 
Peterson et al.\,(2000) and McLure \& Jarvis (2002) the mass estimate is 
reduced by a factor of $\sim 2$ and in this case within the uncertainties in 
better agreement with an empirical upper limit of 
$M_{bh}^{max}\simeq 10^{10} M_\odot$ which also has been suggested by Umemura 
(2003) and Rocca-Volmerange et al.\,(2004)
who studied early phases of galaxy formation and related black hole growth.

For two quasars at redshift $z\simeq 3.5$ we detected quite prominent \ob\
emission. It has been suggested that the emission line profile width of \ob\
can be used as a surrogate of the stellar velocity dispersion $\sigma _\ast$
(Nelson 2000). Applying the relation that connects black hole mass with
stellar velocity dispersion (Tremaine et al.\,2002) the use of \ob\ as
a SMBH mass estimator has been investigated by  Shields et al.\,(2003).
Using the emission line profile width of [\oiii ]$\lambda 5007$ the
estimated the mass of the black holes amount to 
$1.4 \pm 0.3 \times 10^9 M_\odot$ (Q\,0256-0000) and 
$1.5 \pm 0.3 \times 10^9 M_\odot$ (Q\,0302-0019) which are
consistent with the masses provided by \civ , \mgii , and H$\beta$ (Table 3).

While the black hole mass estimates based on \civ $\lambda 1549$ and H$\beta $ 
are quite consistent, the $M_{bh}$ estimate provided by \mgii $\lambda 2798$
indicates lower masses (Table 3).
Although R -- L relations based on statistical results found for quasar 
samples introduce uncertainties of a factor of at least $\sim 3$ in the case of
individual quasars (e.g., Vestergaard 2002) there are some additional 
possibilities which may cause lower $M_{bh}$ estimates based on 
\mgii $\lambda 2798$.

First, the slope $\beta $ of the $R \propto L^\beta$ relationship is still not 
well defined, varying between $\sim 0.5 \,{\rm to}\, 0.7$ 
(e.g., Kaspi et al.\,2000; 
Peterson et al.\,2000; McLure \& Jarvis 2002; Vestergaard 2002; Netzer 2003).
A flatter slope in the $R - L$ relation will result in less massive black
holes. While in the original analysis by Kaspi et al.\,(2000) it is found that
$\beta = 0.70 \pm 0.03$, there are indications that the slope is closer to
$\beta \simeq 0.62\pm0.02$ (Peterson et al.\,2000). McLure \& Jarvis (2002) 
derived $\beta = (0.47 \pm 0.05)$ for the R -- L relation using \mgii . They
determined this flatter slope under the assumption that \mgii\ and H$\beta$
originate at nearly the same BLR radius. Hence, they used H$\beta$ 
reverberation results for Seyfert galaxies and quasars to analyze the
correlation of r$_{BLR}$(\mgii ) with the luminosity 
$\lambda$\,L$_\lambda (3000)$.
However, assuming a slightly flatter slope of $\beta = 0.6$ instead of 
$\beta = 0.7$ will decrease the estimated black hole mass based 
on H$\beta$ by a factor of $\sim 2$. Using furthermore a slightly steeper
slope for \mgii\ (e.g., $\beta = 0.6$ instead of $\beta = 0.47$) will provide 
on average $\sim 2.4 \pm 0.2$ times higher SMBH estimates.
A second aspect as discussed in Sect.\,4.2 is the size of the \mgii\ emitting
region. If \mgii\ originates at slightly larger distances to the center 
of the continuum emission than H$\beta$, the estimated mass can be increased 
by an additional factor. 
     Currently, we suppose that the different SMBH estimates provided by \civ\
     and H$\beta$ compared to those based on \mgii\ are predominantly caused by
     the different slopes of the R-L relations.
Furthermore, it is still an open question whether the R--L relation can be 
extrapolated to quasar luminosities since the relation is derived for active 
galactic nuclei with $L_{bol} \la 10^{46}$ erg\,s$^{-1}$ 
(e.g., Corbett et al.\,2003). In the case of quasars this relation is extended 
to $\sim 10^2$ to $\sim 10^3$ times  more luminous objects.
Yet, in spite of the uncertainties regarding the use of the H$\beta $, \mgii , 
and \civ\ emission line profiles to determine the black hole mass, 
we emphasize that all three emission lines indicate consistently the presence 
of super-massive black holes with masses of 
$10^9  \la M_{bh} \la 10^{10} M_\odot$ at redshifts up to z=5.1.

\subsection{Implications of Black Hole Growth Time Scales}

The presence of SMBHs at high-redshifts and the close relation of quasar
activity and galaxy formation provide valuable constraints on the epoch
when these SMBHs had to start to grow in their forming host galaxies. The
time span that is necessary to build-up a 
$\sim 10^9 \,{\rm to}\, 10^{10} M_\odot$ SMBH can be described by an 
e-folding time for the growth of a single seed black hole (e.g., Haiman \&
Loeb 2001). We use
the following equation to estimate the e-folding time scale and hence
the epoch when the seed black hole started to grow (Salucci et al.\,1999)

\begin{equation}
  M_{bh}(t_{obs}) = M_{bh}^{seed}(t_\circ) \,
                     exp\biggl({{ \eta \tau} \over {\epsilon t_E}}\biggr)
\end{equation}

\noindent
with $\tau = t_{obs} - t_\circ$ as time elapsed from the initial time, $t_o$, 
to the observed time, $t_{obs}$, $M_{bh}^{seed}$ as seed black 
hole mass, $\eta = L_{bol}/L_{edd}$ as Eddington ratio, 
$\epsilon$ as efficiency to convert mass to energy, and $t_E$ as Eddington
time scale, with $t_E = \sigma_T c / 4 \pi G m_p = 3.92 \times 10^8$\,yr
(Rees 1984). The Eddington time $t_E$ describes the time necessary to radiate 
at the Eddington luminosity the entire rest mass of an object.
In hierarchical models for structure formation the first baryonic objects which
collapsed had masses of the order of $\sim 10^4 - 10^6 M_\odot$ (e.g., Silk
\& Rees 1998; Larson 2000; Shibata \& Shapiro 2002; Bromm \& Loeb 2003). This 
provides upper limit estimates for $M_{bh}^{seed}$ of black holes. Recent 
models of early star formation indicate that nearly metal free Pop\,III stars, 
formed at high-redshifts ($z \ga 20$), were predominantly very massive with 
$M\ga 100 M_\odot$ (e.g., Fryer, Woosley, \& Heger 2001; Abel, Bryan, \&
Norman 2002; Bromm, Coppi, \& Larson 2002). 
The stellar remnants are expected to be of the order of $\sim 10 M_\odot$ 
(e.g., Fryer 1999). However, some early star formation models even indicate the
possibility of black hole remnants with several times $\sim 10^3 M_\odot$ 
(Bond, Arnett, \& Carr 1984). In the following the efficiency $\epsilon$ to 
convert mass to energy is assumed to be $\epsilon = 0.1$. 
It should be noted that the efficiency $\epsilon$ can reach values up to 
$\epsilon = 0.42$ for a maximal rotating Kerr black hole.

We employed equation (2) to derive the time $\tau $ which is necessary to 
accumulate the determined mean black hole mass (Table 3) for seed black holes 
with $M_{bh}^{seed} = 10 M_\odot$, $10^3 M_\odot$, and $10^5 M_\odot$,
respectively. 
We find that the time $\tau $ to build-up a SMBH is between several 100 Myr 
and $\sim 1$ Gyr, i.e., almost always $\tau$ is smaller than the age of the 
universe at the corresponding redshift.
To illustrate this in a proper way we used $\tau $ 
to calculate $t_\circ$, i.e. the epoch when the SMBH, observed at $t_{obs}$, 
started to form.
In Figure 4 the time $t_\circ$ is displayed as a function of the observed
redshift $z_{obs}$. 
The determined time scales $\tau $ and hence $t_\circ$ indicate that most
of the SMBH residing in the observed high-redshift quasars started to grow 
at redshifts of $z(t_o) \simeq 5 \,{\rm to}\, 10$.
As can be seen for low massive black hole seeds, i.e., stellar remnants with 
$M_{bh}^{seed} \simeq 10 M_\odot$ some black holes had to have started to grow 
at very early epochs and for Q\,0105$-$2634 and SDSS\,0388$+$0021 the required 
time $\tau $is even larger than the actual cosmic age at $z_{obs}$.
To avoid too long growing times $\tau$, higher $M_{bh}^{seed}$ masses appear 
to be favored. 
Assuming high mass seed black holes with 
$M_{bh}^{seed} = 10^3 \,{\rm to}\, 10^5 M_\odot$, the SMBH in the $z\simeq 3.5$
quasars began to grow at redshifts $z(t_o) \simeq 4$ and earlier. The luminous 
quasars at redshifts $z\ga 4.5$ show a smaller scatter for $t_\circ$ compared 
to those at $z\simeq 3.5$.
Based on our results the SMBH residing in the $z \ga 4.5$ quasars started at 
$z(t_o) \simeq 5 \,{\rm to}\, 10$. 
However, stellar mass black hole seed are still a valid option because a 
flatter slope of the R -- L relation for H$\beta$ with $\beta \simeq 0.6$ 
(Peterson et al.\,2000; McLure \& Jarvis 2002) will yield about two times less 
massive black holes. Hence, the Eddington luminosity
will be reduced and the Eddington ratio will be increased. As a result
the growing time scale $\tau$ will decrease and SMBHs can be built-up with 
stellar mass remnants as black hole seeds as well. 

It is remarkable that most of the super-massive black holes of the studied 
high-redshift quasars are in place at the end of the era of reionization of 
the universe which is supposed to end at $z\simeq 6$ 
(e.g., Becker at al.\,2001; Fan et al.\,2002).
Particularly, for the quasars at redshifts $z\ga 4.5$ we want to emphasize that
the epoch when these black holes started to grow coincides with the beginning
of first intense star formation that marks early phases of the host galaxy 
formation.
This epoch of major star formation also is indicated by the chemical 
enrichment history of the quasar gas. To achieve the estimated metallicities 
of several times solar (e.g., Dietrich et al.\,2003b,c) a major and intense 
star formation episode had to start at $z \simeq 6 \,{\rm to}\, 8$.
Further circumstantial evidence for a vigorous star formation phase at those
high-redshifts is given by the lack of evolution of the iron/$\alpha$ element
ratio compared to results for local quasars (e.g., Barth et al.\,2003; 
Dietrich et al.\,2002,2003a; Freudling et al.\,2003; Iwamuro et al.\,2002; 
Maiolino et al.\,2003; Thompson et al.\,1999).

The presence of super-massive black holes in luminous high-redshift quasars,
the estimated time scale required to build-up black holes of several 
$M_{bh} \simeq 10^9 M_\odot$, the gas metallicity of at least solar up to 
several times solar, and the lack of evolution of the relative iron/$\alpha$
element abundance ratio provide evidence that those quasars reside in already 
mature host galaxies, at least with an evolved massive central component
(Hamann et al.\,2004). 
While the massive galaxy is forming, i.e., major and vigorous star formation 
activity occurs, the super-massive black hole grows. It takes about $\sim 0.5$
Gyr to accumulate a SMBH with several times $M_{bh}\simeq 10^8 M_\odot$ to 
$M_{bh}\simeq 10^9 M_\odot$ that is necessary to power quasar activity.
It takes approximately this same amount of time for a stellar population to 
enrich the gas around quasars to the observed solar or super-solar levels 
(e.g., Dietrich et al. 2003b,c; Matteucci \& Padovani 1993; Matteucci \&
Recchi 2001). 
Therefore, by the time most quasars become visible, even at the highest 
redshifts, a substantially evolved stellar population is harboring
the active nucleus.
This conclusion is supported by some theoretical models of joint
SMBH -- host galaxy evolution (Archibald et al.\,2002, Kawakatu et
al.\,2003). 
Direct imaging studies of low redshift quasars show clearly that the host 
galaxies are characterized by at least moderately old, stellar populations 
on $>$kpc scale (Nolan et al.\,2001; Dunlop et al.\,2003). 
Although there is less imaging data at higher redshifts at least some quasars 
still have substantial hosts (Kukula et al.\,2001).

It can be assumed that early phases of galaxy formation that are accompanied
by intense and vigorous star formation also mark the era when super-massive 
black holes are growing from massive seed remnants (Hamann et al.\,2004).
A joint era of SMBH growth and early phases of galaxy formation can be 
understood in the context of recently suggested galaxy formation models 
(e.g., Archibald et al.\,2002; Di\,Matteo et al.\,2003; 
Granato et al.\,2001,2004). 
These models further suggest that the $M_{bh} \propto \sigma _\ast ^4$ relation
evolves over cosmic time scales and is finally established in the local
universe (Di\,Matteo et al.\,2003). 
A less steep $M_{bh} \propto \sigma _\ast$ relation also has been suggested
by Shields et al.\,(2003). This may provide an explanation to avoid
huge host galaxy masses ($\sim 10^{13} M_\odot$) and large stellar velocity
dispersions ($\sigma _\ast \simeq 700$ km\,s$^{-1}$) which have never been
observed (e.g., Netzer 2003).

\section{Conclusion}
We have analyzed a sample of 15 luminous high-redshift quasars 
($3.3 \la z \la 5.1$) to estimate the mass of super-massive black holes. 
For the first time results for $M_{bh}$ based on three different emission 
lines, \civ , \mgii , and H$\beta$, are compared for high-redshift quasars.

We determine black hole masses in the range of $M_{bh}\simeq 2\times 10^8 
M_\sun$ up to $M_{bh}\simeq 4 \times 10^{10} M_\sun$  
Similar to prior studies there appears to be an approximate upper limit for 
$M_{bh}$ at $\sim 10^{10} M_\odot$.
The black hole mass estimates based on \civ\ and H$\beta $ are quite consistent
while \mgii\ indicates by a factor of $\sim 5$ lower SMBH masses. 
In spite of the uncertainties regarding the use of the H$\beta $, \mgii , 
and \civ\ emission line profiles to determine the black hole mass, all three 
emission lines indicate consistently the presence of super-massive black holes 
with masses on average $\sim 6 \times 10^9 M_\odot$ at redshifts up to z=5.1.

We determine an Eddington luminosity ratio of $L_{bol} / L_{edd}$ in the range 
of $\sim 0.5$ to $\sim 7$. This is consistent with results of studies of 
quasars sample that are dominated by low and intermediate redshift quasars. 
This indicates that high-redshift quasars are not different regarding the 
Eddington ratio $L_{bol} / L_{edd}$ compared to those at lower redshift.

To date the epoch when the SMBHs which drive high-redshift quasar activity 
should have started to grow, we assume logarithmic growth by spherical 
accretion with a mass to energy conversion efficiency of $\epsilon = 0.1$. 
Generally, the growth-time $\tau$ is of the order of several $\sim 100$ Myr
Assuming higher mass seed black holes ($M_{bh}^{seed} = 10^3 \,{\rm to}\, 
10^5 M_\odot$) the SMBH in the $z\simeq 3.5$ quasars begun to grow at redshifts
$z(t_o) \simeq 4$ and earlier while the quasars with $z\ga 4.5$ started at 
$z(t_o) \simeq 5 \,{\rm to}\, 10$. 

The presence of super-massive black holes in luminous high-redshift quasars,
the estimated time scale required to build-up black holes of several times
$M_{bh} \simeq 10^9 M_\odot$, the several times solar gas metallicity and 
the lack of evolution of the relative iron/$\alpha$ element abundance ratio 
provide evidence that luminous quasars at high-redshift reside in already 
mature host galaxies, at least with an substantially evolved massive central 
stellar component. 
This result indicates that early phases of massive galaxy formation that are 
accompanied by intense and vigorous star formation also mark the era when 
super-massive black holes are growing from massive seed remnants.  
The estimated black hole growth time scale suggest a delay of about $\sim 0.5$ 
Gyr of the quasar activity with respect to preceding intense star formation 
which is consistent with the observed high gas metallicity.

\begin{acknowledgements}
      We dedicate this Paper to the memory of Richard J. Elston, who left us
      far too early on January 26, 2004.

      MD and FH acknowledge support from NASA grant NAG 5-3234 and NSF grant 
      AST-99-84040 (University of Florida).
\end{acknowledgements}

\appendix

\section{Appendix}

The equations we applied, given by Kaspi et al.\,(2000), McLure \& Jarvis 
(2002), Vestergaard (2002), and Warner et al.\,(2003) have been modified for 
the use of the continuum luminosity $L_\lambda$ that we measured at 
$\lambda = 1450$\,\AA\ for each of the high-redshift quasars of this study. 
We explicitly present the dependence on the spectral index $\alpha$ which 
transforms the continuum luminosity measured at different wavelengths. 
The uniform use of $L_\lambda (1450 {\rm \AA })$ does not introduce an 
additional uncertainty, because instead of assuming a fixed continuum slope 
$\alpha$ to rescale $L_\lambda $ as in prior studies we used the measured 
continuum slope $\alpha $ for each quasar which we determined for the power-law
continuum fit ($F_\nu \propto \nu ^\alpha$, Table 2) in the multi-component 
analysis of the quasar spectra (Dietrich et al.\,2002a,2003a). The continuum
slope is measured for a wavelength range of $\lambda \lambda \sim 1100 
\,\,{\rm to}\,\,5500$ \AA\ ($z=3.5$ quasars) and $\lambda \lambda \sim 1100 
\,\,{\rm to}\,\, 4300$ \AA\ ($z=4.5$ quasars).

For the H$\beta$ emission line profile we determined $M_{bh}$ applying
the equation given by Kaspi et al.\,(2000) (eq.\,A1) and presented by
McLure \& Jarvis (2002) (eq.\,A2). The main difference is given by the slope 
of the radius -- luminosity (R -- L) relation.

\begin{equation}
M_{bh} = 4.82 \times 10^6 \times 
         \biggl({1450 \over 5100}\biggr)^{0.7(1+\alpha)} \times
         \biggl({\lambda L_\lambda (1450 {\rm \AA}) \over 
                 {10^{44} \,{\rm erg/s}}}\biggr)^{0.7} 
         \times \biggl({{\rm FWHM}(H\beta) \over {10^3 \,{\rm km/s}}}\biggr)^2 
         \quad M_\odot
\end{equation}

\begin{equation}
M_{bh} = 4.74 \times 10^6 \times 
         \biggl({1450 \over 5100}\biggr)^{0.61(1+\alpha)} \times
         \biggl({\lambda L_\lambda (1450 {\rm \AA}) \over 
                 {10^{44} \,{\rm erg/s}}}\biggr)^{0.61} 
         \times \biggl({{\rm FWHM}(H\beta) \over {10^3 \,{\rm km/s}}}\biggr)^2 
         \quad M_\odot
\end{equation}

The black hole mass is estimated based on the \civ $\lambda 1549$ line profile
width applying the following equations (Vestergaard 2002, eq.\,A3; 
Warner et al.\,2003, eq.\,A4). Warner et al.\,(2003) assumed that the radius
of the \civ\ BLR is half the size of the H$\beta $ emitting region which is
well justified by variability studies (Peterson 1993).

\begin{equation}
M_{bh} = 1.58 \times 10^6 \times 
         \biggl({1450 \over 1350}\biggr)^{0.7(1+\alpha)} \times
         \biggl({\lambda L_\lambda (1450 {\rm \AA}) \over 
                 {10^{44} \,{\rm erg/s}}}\biggr)^{0.7} 
         \times \biggl({{\rm FWHM(CIV)} \over {10^3 \,{\rm km/s}}}\biggr)^2
         \quad M_\odot
\end{equation}

\begin{equation}
M_{bh} = 2.41 \times 10^6 \times 
         \biggl({1450 \over 5100}\biggr)^{0.7(1+\alpha)} \times
         \biggl({\lambda L_\lambda (1450 {\rm \AA}) \over 
                {10^{44} \,{\rm erg/s}}}\biggr)^{0.7} 
         \times \biggl({{\rm FWHM(CIV)} \over {10^3 \,{\rm km/s}}}\biggr)^2
         \quad M_\odot
\end{equation}

To estimate $M_{bh}$ based on the \mgii $\lambda 2798$ emission line profile
width we employed the relation given by McLure \& Jarvis (2002).

\begin{equation}
M_{bh} = 3.37 \times 10^6 \times 
         \biggl({1450 \over 3000}\biggr)^{0.47(1+\alpha)} \times
         \biggl({\lambda L_\lambda (1450 {\rm \AA}) \over 
                 {10^{44} \,{\rm erg/s}}}\biggr)^{0.47} 
         \times \biggl({{\rm FWHM(MgII)} \over {10^3 \,{\rm km/s}}}\biggr)^2
         \quad M_\odot
\end{equation}

\clearpage

\figcaption[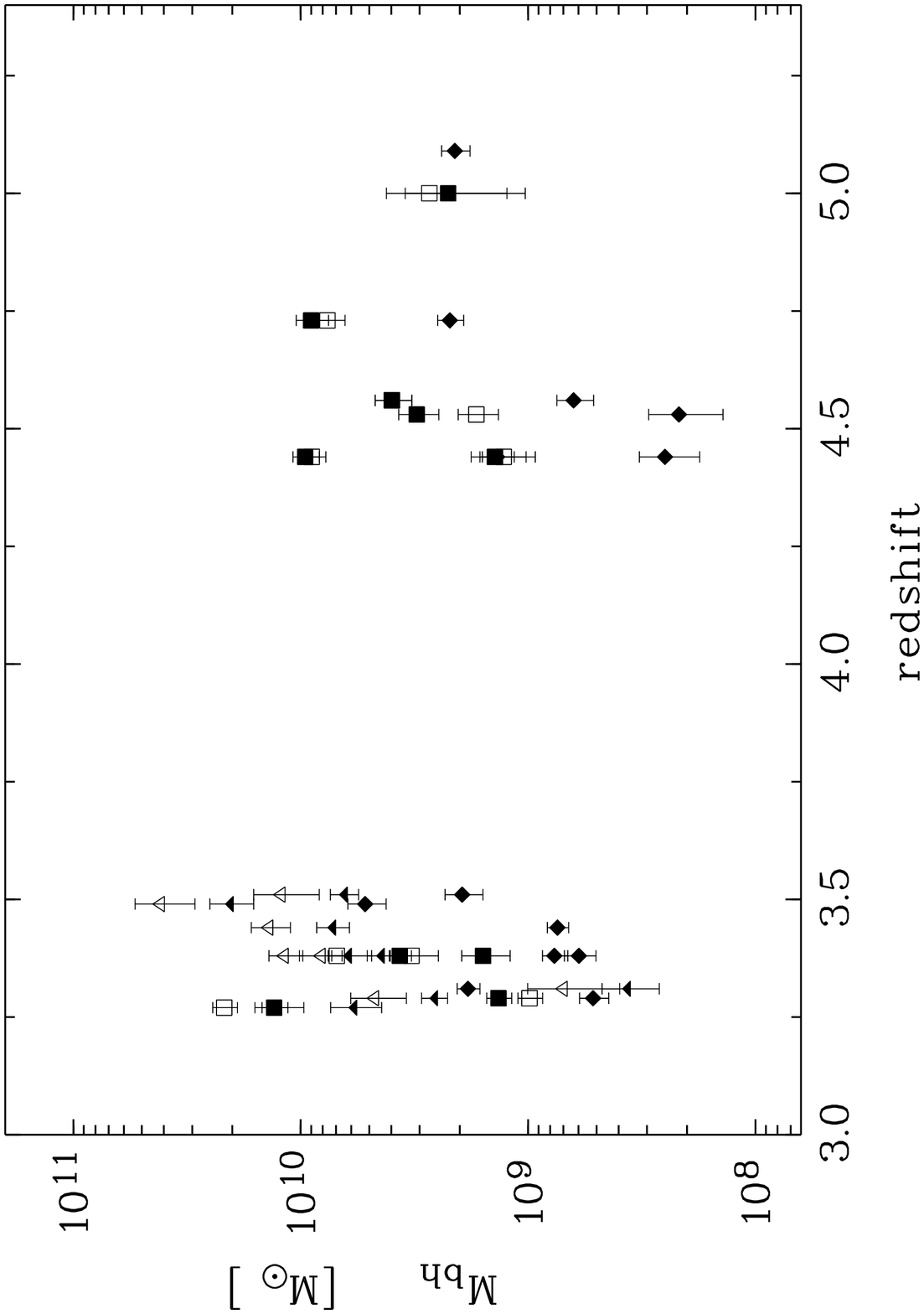]{Estimates SMBH masses for the high-redshift
            quasars as a function of redshift using eq.\,(A1) to (A5). The 
            triangles indicate 
            $M_{bh}$ based on H$\beta$ ($\triangle$ -- Kaspi et al.\,2000;
            $\blacktriangle$ -- McLure \& Jarvis 2002), the squared symbols
            show $M_{bh}$ based on \civ\ ($\square $ -- Warner et al.\,2003; 
            $\blacksquare $ -- Vestergaard 2002), and the filled diamonds
            display $M_{bh}$ based on \mgii\ ($\blacklozenge $ 
            McLure \& Jarvis 2002).}

\figcaption[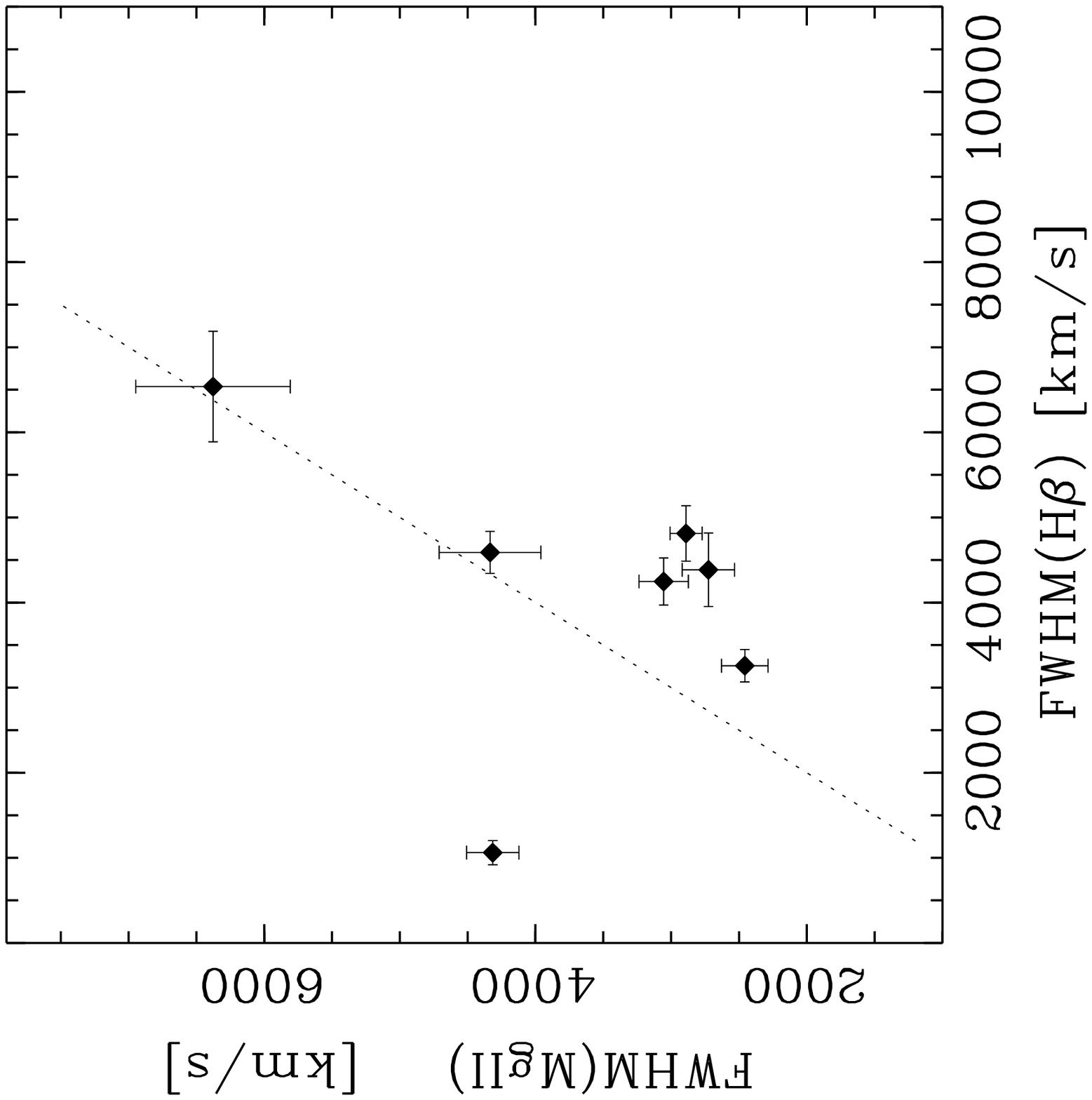]{Comparison of the line profile width, FWHM, of 
            \mgii $\lambda 2798$ and H$\beta $ for the high-redshift 
            quasars at $z\simeq 3.5$. The dotted line indicates a perfect
            one-to-one relation.}

\figcaption[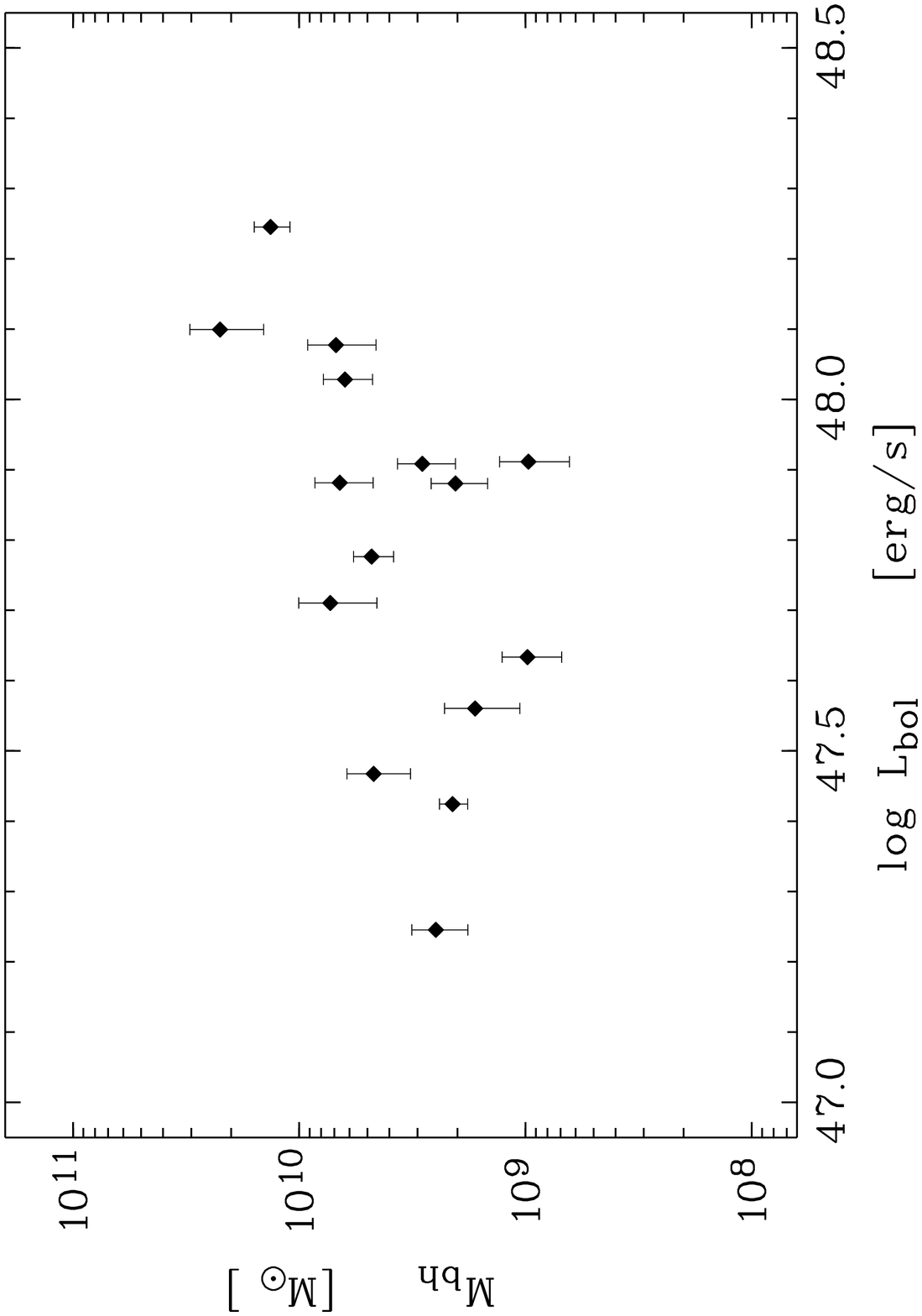]{Mean SMBH masses for the high-redshift
            quasars as a function of bolometric luminosity, $L_{bol}$, are
            shown. The displayed errors are taking into account the
            measurement uncertainties and the distribution of the individual
            SMBH mass estimates in quadrature (Table 3).}

\figcaption[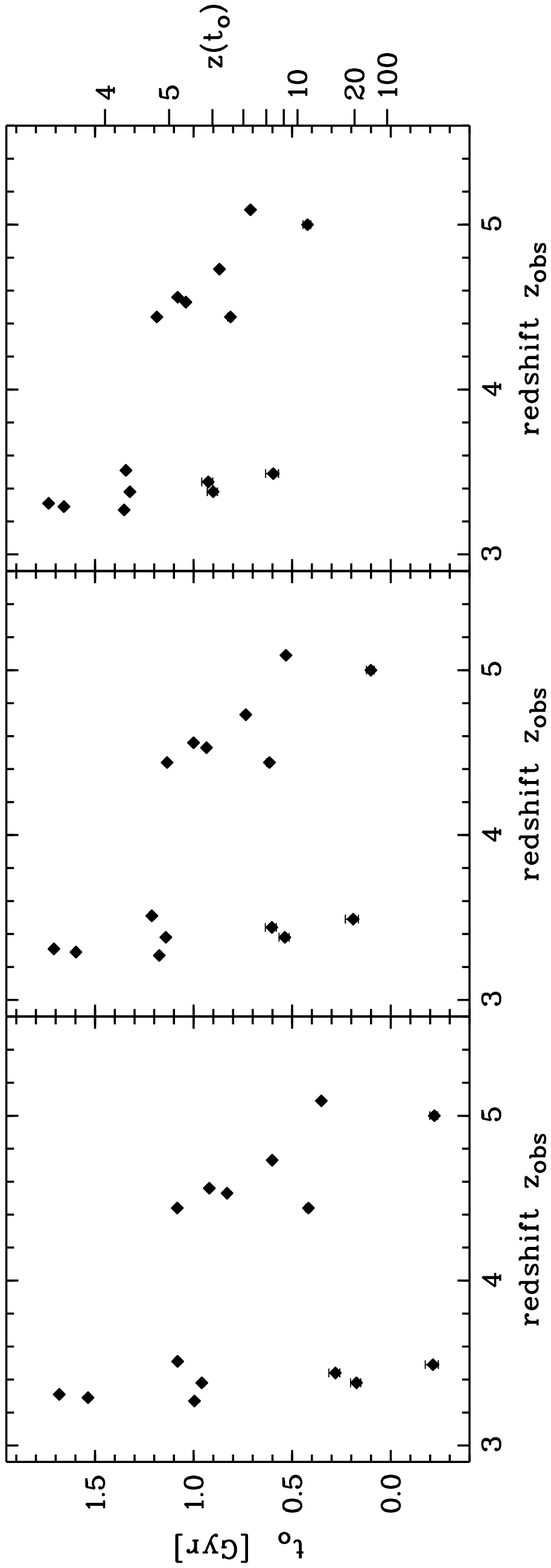]{The time $t_o = \tau _{obs} - \tau$  is 
            plotted as a function of redshift for several seed black hole
            masses to illustrate when the super-massive black holes had had to
            start to form.
            The three panels display results for different seed black hole 
            masses with
            $M_{bh}^{seed} = 10 M_\odot$ (left), 
            $M_{bh}^{seed} = 10^3 M_\odot$ (center), and
            $M_{bh}^{seed} = 10^5 M_\odot$ (right).
            The redshift which corresponds to $t_o$ is labeled at the right
            as $z(t_o)$
            (${\rm H}_o = 65$ km\,s$^{-1}$\,Mpc$^{-1}$, 
            $\Omega _M = 0.3$, $\Omega _\Lambda = 0.0$).} 

%

\begin{figure}
\plotone{f1.eps}
\label{fig1}
\end{figure}

\begin{figure}
\plotone{f2.eps}
\label{fig2}
\end{figure}

\begin{figure}
\plotone{f3.eps}
\label{fig3}
\end{figure}

\begin{figure}[t]
\epsscale{0.50}
\plotone{f4.eps}
\label{fig4}
\end{figure}

\hspace*{-15mm}
\begin{deluxetable}{lcccccccc}
\tablewidth{0pt}
\tabletypesize{\scriptsize}
\tablecaption{The high-redshift quasars have been observed at the observatories
              and dates as given in (2) and (3). In addition, the achieved 
              spectral resolution for the emission line profiles of H$\beta$, 
              \mgii ,
              and \civ\ is listed in unit of km\,s$^{-1}$ (columns 4 to 6). 
              The redshift of the quasars is given in (7). The rest frame 
              continuum luminosity $\lambda {\rm L}_\lambda $(1450\AA) and 
              the bolometric luminosity, ${\rm L}_{bol}$, are listed in (8)
              and (9).}
\tablehead{
\colhead{quasar} &
\colhead{site} &
\colhead{date} &
\colhead{ } &
\colhead{resol.} &
\colhead{ } &
\colhead{redshift} &
\colhead{log $\lambda L_\lambda (1450 {\rm \AA})$} &
\colhead{$log L_{bol}$}\\
\colhead{ } &
\colhead{ } &
\colhead{ } &
\colhead{CIV} &
\colhead{MgII } &
\colhead{H$\beta$} &
\colhead{ } &
\colhead{ } &
\colhead{ } \\
\colhead{ } &
\colhead{ } &
\colhead{ } &
\colhead{ } &
\colhead{[km\,s$^{-1}$]} &
\colhead{ } &
\colhead{ } &
\colhead{[erg\,s$^{-1}$]} &
\colhead{[erg\,s$^{-1}$]}\\
\colhead{(1)} &
\colhead{(2)} &
\colhead{(3)} &
\colhead{(4)} &
\colhead{(5)} &
\colhead{(6)} &
\colhead{(7)} &
\colhead{(8)} &
\colhead{(9)} 
}
\startdata
BRI 0019-1522 &CTIO&09-00&300     &230     &\nodata &4.53& 46.92&47.56\\
BR 0103+0032  &CTIO&09-00&300     &230     &\nodata &4.44& 46.99&47.63\\
Q 0103-260    &VLT &12-98&750     &\nodata &\nodata &3.38& 46.83&47.47\\
              &NTT &10-99&\nodata &500     &430     &    &      &     \\
Q 0105-2634   &NTT &10-99&\nodata &500     &430     &3.49& 47.46&48.10\\
PSS J0248+1802&CTIO&09-00&300     &230     &\nodata &4.44& 47.24&47.88\\
Q 0256-0000   &NTT &10-99&300     &500     &430     &3.38& 47.14&47.78\\
Q 0302-0019   &NTT &09-99&300     &500     &430     &3.29& 47.24&47.88\\
SDSS 0338+0021&VLT &12-98&750     &\nodata &\nodata &5.00& 46.61&47.25\\
PC 1158+4635  &KECK&05-00&300     &520     &\nodata &4.73& 47.39&48.03\\
SDSS 1204-0021&VLT &04-01&\nodata &525     &\nodata &5.09& 46.79&47.42\\
Q 2050-259    &CTIO&09-00&\nodata &230     &230     &3.51& 47.44&48.08\\
PKS 2126-158  &CA  &08-93&300     &\nodata &\nodata &3.27& 47.61&48.25\\
              &CTIO&09-00&\nodata &230     &230     &    &      &     \\
Q 2227-3928   &NTT &10-99&\nodata &500     &430     &3.44& 47.07&47.71\\
BRI 2237-0607 &CTIO&09-00&300     &230     &\nodata &4.56& 47.27&47.91\\
Q 2348-4025   &NTT &10-99&\nodata &500     &430     &3.31& 47.27&47.91\\
\enddata
\end{deluxetable}

\hspace*{-15mm}
\begin{deluxetable}{lcccc}
\tablewidth{0pt}
\tabletypesize{\scriptsize}
\tablecaption{Measurements of the emission line profile widths (FWHM) for
         H$\beta$, \mgii $\lambda 2798$, and \civ $\lambda 1549$ are given for 
         the studied high-redshift quasar sample. 
         The slope $\alpha$ of the power-law continuum fit 
         ($F_\nu \propto \nu ^\alpha$) given in column (5).
}
\tablehead{
\colhead{quasar} &
\colhead{FWHM(H$\beta$)} &
\colhead{FWHM(\mgii )} &
\colhead{FWHM(\civ )} &
\colhead{$\alpha$}\\
\colhead{ } &
\colhead{ } &
\colhead{[km\,s$^{-1}$] } &
\colhead{ } &
\colhead{ }\\
\colhead{(1)} &
\colhead{(2)} &
\colhead{(3)} &
\colhead{(4)} &
\colhead{(5)} 
}
\startdata
BRI 0019-1522 &\nodata     &$1993\pm353$&$4078\pm347$&$+0.10\pm 0.09$\\
BR 0103+0032  &\nodata     &$1869\pm268$&$2626\pm118$&$-0.46\pm 0.03$\\
Q 0103-260    &$4384\pm432$&$2725\pm193$&$3257\pm350$&$-1.33\pm 0.06$\\
Q 0105-2634   &$6537\pm650$&$6377\pm570$&\nodata     &$-0.69\pm 0.06$\\
PSS J0248+1802&\nodata     &$3811\pm279$&$5620\pm224$&$-0.48\pm 0.08$\\
Q 0256-0000   &$4247\pm277$&$3055\pm182$&$3784\pm163$&$-0.41\pm 0.05$\\
Q 0302-0019   &$3257\pm191$&$2457\pm171$&$2102\pm108$&$-0.21\pm 0.03$\\
SDSS 0338+0021&\nodata     &\nodata    &$4500\pm1200$&$-0.01\pm 0.20$\\
PC 1158+4635  &\nodata     &$4559\pm257$&$4833\pm300$&$-0.37\pm 0.05$\\
SDSS 1204-0021&\nodata     &$6090\pm380$&\nodata     &$-0.43\pm 0.06$\\
Q 2050-259    &$4589\pm247$&$4335\pm375$&\nodata     &$-0.15\pm 0.18$\\
PKS 2126-158  &$2707\pm320$&\nodata     &$4991\pm176$&$-1.09\pm 0.09$\\
Q 2227-3928   &$4812\pm326$&$2890\pm118$&\nodata     &$-0.85\pm 0.05$\\
BRI 2237-0607 &\nodata     &$2517\pm214$&$3550\pm269$&$-0.55\pm 0.02$\\
Q 2348-4025   &$1061\pm142$&$4315\pm193$&\nodata     &$-0.53\pm 0.07$\\
\enddata
\end{deluxetable}

\hspace*{-15mm}
\begin{deluxetable}{lccccccc}
\tablewidth{0pt}
\tabletypesize{\scriptsize}
\tablecaption{The SMBH masses estimates for the studied high-redshift quasar 
              sample. The derived black hole masses based on H$\beta$ using  
              Kaspi et al.\,(2000, K00) and McLure \& Jarvis (2002, MJ02) 
              are given in (2) and (3). The SMBH estimates employing
              \civ $\lambda 1549$ are listed in column (4) and (5) (
              Vestergaard 2002, MV02; Warner et al\,2003, CW03) and using 
              \mgii $\lambda 2798$ in (6) (McLure \& Jarvis 2002, MJ02).
              In column (7) the mean super-massive black hole mass is listed
              and the resulting Eddington ratio ${\rm L}_{bol}/{\rm L}_{edd}$ 
              is given in column (8).}
\tablehead{
\colhead{quasar} &
\colhead{H$\beta$ (K00) } &
\colhead{H$\beta$ (MJ02)} &
\colhead{\civ\ (MV02)   } &
\colhead{\civ\ (CW03)   } &
\colhead{\mgii\ (MJ02)  } &
\colhead{${\rm M}_{bh}$ } &
\colhead{${\rm L}_{bol}/{\rm L}_{edd}$ } \\
\colhead{ } &
\colhead{ } &
\colhead{ } &
\colhead{M$_{bh}$ [$10^9 {\rm M}_\odot$] } &
\colhead{ } &
\colhead{ } &
\colhead{[$10^9 {\rm M}_\odot$] } &
\colhead{ }\\
\colhead{(1)} &
\colhead{(2)} &
\colhead{(3)} &
\colhead{(4)} &
\colhead{(5)} &
\colhead{(6)} &
\colhead{(7)} & 
\colhead{(8)} 
}
\startdata
BRI 0019-1522 & \nodata     & \nodata     &$ 3.1\pm 0.6$&$ 1.7\pm 0.3$&$0.22\pm 0.08$&$ 1.7\pm0.6$&1.73\\
BR 0103+0032  & \nodata     & \nodata     &$ 1.4\pm 0.4$&$ 1.3\pm 0.4$&$0.25\pm 0.07$&$ 1.0\pm0.3$&3.48\\
Q 0103-260    &$11.8\pm 2.0$&$ 6.2\pm 1.3$&$ 1.6\pm 0.4$&$ 3.3\pm 0.8$&$0.60\pm 0.10$&$ 4.7\pm1.5$&0.50\\
Q 0105-2634   &$41.4\pm12.2$&$20.6\pm 4.5$& \nodata     & \nodata     &$5.2\pm 1.0$  &$22.4\pm8.0$&0.45\\
PSS J0248+1802& \nodata     & \nodata     &$ 9.6\pm 1.3$&$ 8.9\pm 1.2$&$1.4\pm 0.2$  &$ 6.6\pm1.9$&0.91\\
Q 0256-0000   &$ 8.1\pm 2.0$&$ 4.5\pm 0.6$&$ 3.7\pm 1.4$&$ 6.9\pm 0.4$&$0.77\pm 0.10$&$ 4.8\pm0.9$&0.99\\
Q 0302-0019   &$ 4.7\pm 1.3$&$ 2.6\pm 0.3$&$ 1.4\pm 0.2$&$ 1.0\pm 0.1$&$0.52\pm 0.08$&$ 2.0\pm0.6$&2.95\\
SDSS 0338+0021& \nodata     & \nodata     &$ 2.3\pm 1.2$&$ 2.7\pm 1.5$& \nodata      &$ 2.5\pm0.7$&0.56\\
PC 1158+4635  & \nodata     & \nodata     &$ 9.0\pm 1.5$&$ 7.6\pm 1.2$&$2.2\pm 0.3$  &$ 6.3\pm1.5$&1.35\\
SDSS 1204-0021& \nodata     & \nodata     & \nodata     & \nodata     &$2.1\pm 0.3$  &$ 2.1\pm0.3$&1.00\\
Q 2050-259    &$12.2\pm 3.9$&$ 6.5\pm 0.9$& \nodata     & \nodata     &$2.0\pm 0.4$  &$ 6.9\pm2.3$&1.38\\
PKS 2126-158  &$12.8\pm 3.1$&$ 5.9\pm 1.5$&$13.1\pm 1.7$&$21.7\pm 2.7$& \nodata      &$13.4\pm2.4$&1.01\\
Q 2227-3928   &$13.8\pm 2.7$&$ 7.3\pm 1.2$& \nodata     & \nodata     &$0.74\pm 0.08$&$ 7.3\pm1.5$&0.56\\
BRI 2237-0607 & \nodata     & \nodata     &$ 4.0\pm 0.7$&$ 4.0\pm 0.7$&$0.63\pm 0.12$&$ 2.9\pm0.8$&2.25\\
Q 2348-4025   &$0.70\pm0.30$&$0.37\pm0.10$& \nodata     & \nodata     &$1.8\pm 0.2$  &$ 1.0\pm0.3$&6.67\\
\enddata
\end{deluxetable}

\end{document}